\documentclass[preprint]{aastex}

\usepackage{amsmath}
\usepackage{color}
\usepackage[dvipdfmx]{hyperref}

\newcommand{\lsim}{\protect\raisebox{-0.8ex}{$\:\stackrel{\textstyle <}{\sim}\:$}} 
\newcommand{\gsim}{\protect\raisebox{-0.8ex}{$\:\stackrel{\textstyle >}{\sim}\:$}}

\shorttitle{Constraints on neutrino masses from the lensing dispersion of Type Ia supernovae}
\shortauthors{Ryuichiro Hada and Toshifumi Futamase}

\begin{document}

\title{Constraints on neutrino masses from the lensing dispersion \\ of Type Ia supernovae}

\author{Ryuichiro Hada\altaffilmark{1,2} and Toshifumi Futamase\altaffilmark{3}}

\altaffiltext{1}{Astronomical Institute, Tohoku University, Aoba-ku, Sendai 980-8578, Japan}

\altaffiltext{2}{Division for Interdisciplinary Advanced Research and Education, Tohoku University, Aoba-ku, Sendai 980-8578, Japan }

\altaffiltext{3}{Department of Astrophysics and Meteorology, Kyoto Sangyo University, Kita-ku, Kyoto 603-8555, Japan}

\email{r.hada@astr.tohoku.ac.jp; tof@cc.kyoto-su.ac.jp}

\begin{abstract}

We investigate how accurately the total mass of neutrinos is constrained from the magnitude dispersion of SNe Ia due to the effects of gravitational lensing. For this purpose, we use the propagation equation of light bundles in a realistic inhomogeneous universe and propose a sample selection for supernovae to avoid difficulties associated with small-scale effects such as strong lensing or shear effects. With a fitting formula for the nonlinear matter power spectrum taking account of the effects of massive neutrinos, we find that in our model it is possible to obtain the upper limit $\Sigma m_{\nu} \simeq 1.0[{\rm eV}]$ for future optical imaging surveys with the {\it Wide-Field InfraRed Survey Telescope} and Large Synoptic Survey Telescope. Furthermore, we discuss how far we need to observe SNe Ia and to what extent we have to  reduce the magnitude error except for lensing in order to realize the current tightest limit $\Sigma m_{\nu} < 0.2[{\rm eV}]$.

\end{abstract}


\section{\label{sec1}Introduction}

The detection of neutrino oscillation tells us that neutrinos have finite masses.   
However, it tells us only the difference of the squared masses between the neutrino mass eigenstates and  thus we can only obtain the lower limit of the total mass, 0.056(0.096)[eV] $ < \Sigma m_{\nu}$ in the normal (inverted) hierarchy~\citep{2006PhR...429..307L}. 
The knowledge of mass for each neutrino flavor 
will be the important information beyond the standard model of elementary particles.
Therefore, measuring the absolute mass scale, that is, constraining the upper limit of the total neutrino mass is one of the most important tasks in current physics. 

Neutrinos, which are much lighter than  cold dark matter (CDM), move around freely to smooth out the density perturbation in small-scales, and therefore have significant effects on  the cosmological structure formation. Thus, the observation of a matter power spectrum 
is expected to place a certain constraint on the neutrino mass scale. In fact, there are tight constraints on the total neutrino mass obtained from cosmological observations: $\Sigma m_{\nu} < 0.23\ [{\rm eV}]$(95\%CL) from the cosmic microwave background (CMB) anisotropy combined with the distance measurements from the baryon acoustic oscillation (BAO)~\citep{2014A&A...571A..16P} or $\Sigma m_{\nu} < 0.17\ [{\rm eV}]$(95\%CL) from combining the Ly$\alpha$ forest with CMB, supernovae, and galaxy clustering constraints~\citep{2006JCAP...10..014S}.

In this paper, we consider to constrain the mass of neutrinos using the magnitude dispersion of Type Ia supernovae (SNe Ia) due to the effects of weak gravitational lensing. The distance-redshift relation from observations of distant SNe Ia, which is known as cosmological standard candles, has been used to determine the cosmological parameters; in particular, it shows the accelerated  expansion of the present universe~\citep{1998AJ....116.1009R,1999ApJ...517..565P}. 
This observation is based on the fact that there is a one-to-one correspondence between the magnitude and redshift for SNe Ia. However, the magnitude of SNe at the same redshift actually has some dispersion caused by various factors. This magnitude dispersion of SNe Ia is divided into the following two parts: (1) constant error, which includes the intrinsic luminosity dispersion or the uncertainty due to light-curve fitting, and (2) lensing dispersion, which increases with redshift. The lensing dispersion is, in a sense, a type of systematic uncertainty; however, it results from the weak lensing caused by the large scale structure (LSS) as well as cosmic shear, and then provides us with the information about the LSS. In fact, this signal is verified, albeit only at around $2 \sigma$ level, by some SNe observations~\citep{2010MNRAS.405..535J,2010A&A...514A..44K,2014ApJ...780...24S}.   

The lensing dispersion for the magnitude of SNe Ia has been studied previously. Most studies focused on the magnitude probability distribution function (PDF) and some researchers
have assumed a universal magnitude PDF and calibrated the coefficients, e.g., using {$N$}-body simulations, or considering a model of the universe and directly computing the magnitude PDF by ray-shooting simulations~\citep[][and references therein]{2013PhRvD..88f3004M}. Moreover, some authors have investigated, in particular, small-scale structure using the lensing effects for SNe Ia~\citep{1991ApJ...374...83R,1999MNRAS.305..746M,1999ApJ...519L...1M,1999A&A...351L..10S,2014MNRAS.442.2659F,2014MNRAS.443L...6C,2016MNRAS.455..552B}. 

In this paper we study the distance-redshift relation in an inhomogeneous universe~\citep{1987MNRAS.228..653S,1989PhRvD..40.2502F}, which is derived from first principles without any models or parameters in the framework of geometrical optics, to estimate the (de)magnificatin of SNe Ia. As we will see later, we  use the weak lensing approximation in order to associate the magnitude dispersion with the matter power spectrum. Therefore, we discuss a method of sample selection for choosing SNe Ia 
in order to avoid any complications associated with small-scale structures (e.g. the shear effects or strong lensing). This selection corresponds to setting an upper limit for the wave number in the context of the power spectrum. Accordingly, we considered, in our previous work~\citep{2014JCAP...12..042H}, we considered a model in which the wave numbers are universally (at each redshift) cut off at $k = 1[h{\rm Mpc}^{-1}]$. In this paper, we more realistically set upper limits of the wave number at each  redshift using the Press-Schechter model. Furthermore, we discuss  the effect of massive neutrinos on the lensing dispersion in the distance-redshift relation , and then forecast to what extent we will constrain neutrino masses from observations of SNe Ia 
in the planned surveys such as those with the {\it Wide-Field InfraRed Survey Telescope}~\citep[{\it WFIRST};][]{2015arXiv150303757S} and  the Large Synoptic Survey Telescope~\citep[LSST;][]{2008arXiv0805.2366I,2009arXiv0912.0201L}.

The paper is organized as follows. In Sec.\ref{sec2}, we first introduce the relation between the magnification (or demagnification) of SNe Ia and the matter density contrast along a line of sight, and then obtain the expression of variance of the apparent magnitude PDF of SNe Ia due to lensing. In addition, we discuss about a sample selection of SNe Ia to address some difficulties caused by small-scale structures, and define a critical value of the wave number in order to connect the selection of SNe to the theoretical formulation for the lensing dispersion. In Sec.\ref{sec3}, we calculate the lensing dispersion for $\Lambda$CDM models with massive neutrinos and forecast constraints on the parameter $\Sigma m_{\nu}$ from some future surveys using the Fisher information matrix. Finally, Sec.\ref{sec4} is devoted to a summary and discussion.

\section{\label{sec2}Lensing dispersion of SNe Ia}

In this section, we show how to estimate the variance of the apparent magnitude PDF for SNe Ia due to lensing. To this end, we introduce the relation between the magnification (or demagnification) and the matter density contrast along a line of sight in Sec.~\ref{sec2_A} and consider a selection for SNe Ia to overcome some difficulties caused by small-scale structures in Sec.~\ref{sec2_B}. Moreover, in Sec.~\ref{sec2_C}, we define a critical value of the wave number to connect the selection to the theoretical formulation.

\subsection{\label{sec2_A}Formalism}

Assuming a flat universe ($\Omega_{k0} = 0$), the luminosity distance in a homogeneous FRW universe is defined by $d^{\rm{FRW}}_{\rm{L}}(z)=(1+z)\chi(z)$, where $\chi(z)$ is the comoving distance, and the magnitude-redshift relation is then described as follows:
\begin{eqnarray}
	m(z)
		&=&5\log_{10} d^{\rm{FRW}}_{\rm{L}}(z)+M, \label{eq.m-db}
\end{eqnarray}
where $M$ is the absolute magnitude. However, in a realistic inhomogeneous universe, 
the luminosity distance for a source at $z=z_{s}$ is corrected by~\citep{1987MNRAS.228..653S,1989PhRvD..40.2502F,2009PThPh.122..511O},
\begin{eqnarray}
	\delta_{d}(z_s,\hat{\bf{n}})
  		&\equiv& \frac{d_{\rm{L}}(z_s,\hat{\bf{n}}) - d^{\rm{FRW}}_{\rm{L}}(z_s)}{d^{\rm{FRW}}_{\rm{L}}(z_s)}
  		\nonumber \\
		&=&{\bf{v}}_s \cdot \hat{\bf{n}}-\frac{1}{\chi_s}\left[\frac{1}{aH}\right]_s({\bf{v}}_s \cdot \hat{\bf{n}}-{\bf{v}}_o \cdot \hat{\bf{n}}) 
		\nonumber \\
		& &-\int_{0}^{\chi_{s}} d\chi\frac{(\chi_s-\chi)\chi}{\chi_s}\left(\Delta \Psi (z,\hat{\bf{n}})+\tilde{\sigma}^2\right). \label{eq.d-z}
\end{eqnarray}
In the second line, $\hat{\bf{n}}$ is the source direction, $\chi_s \equiv \chi(z_s)$ is the source comoving distance, and ${\bf{v}}_s$ and  ${\bf{v}}_o$ are the source and observer peculiar velocities, respectively, which describe the Doppler effects. The third line corresponds to the convergence of the bundle of light rays ($\Psi$ is the Newtonian potential generated by the density inhomogeneity) and, particularly, $\tilde{\sigma}^2$ represents the (squared) shear.  

Then, the change in the apparent magnitude due to lensing is written as follows:
\begin{eqnarray}
	\delta m_{\rm lens}(z_s,\hat{\bf{n}})
		&=&\frac{5}{\ln10} \ln (1+\delta_{d}(z_s,\hat{\bf{n}})) 
		\nonumber \\
		&\simeq& \frac{5}{\ln10}\delta_{d}(z_s,\hat{\bf{n}})
		\nonumber \\
		&\simeq& -\frac{15H_0^2\Omega_{m0}}{2\ln10}\int_{0}^{\chi_{s}} d\chi\frac{(\chi_s-\chi)\chi}{\chi_s}(1+z)\delta_m(z,\hat{\bf{n}}), \label{eq.m-z}
\end{eqnarray}
where $H_{0}$ is the present Hubble parameter, $\Omega_{m0}$ is the present matter density parameter, and $\delta_m$ is the relative perturbation of matter. Here, we ignored the second or upper order terms of $\delta_{d}$ in the second equality because the correction of luminosity distance, Eq.~(\ref{eq.d-z}), is derived based on the estimation that the net magnification (or demagnification) is so small that the linear approximation is valid. Moreover, when writing the third  equality, we used the Poisson equation,
\begin{eqnarray}
	\Delta \Psi = 4 \pi G a^2 \delta \rho_m = \frac{3 H_0^2\Omega_{m0}}{2}\frac{\delta_m}{a}, \label{eq.poisson} 
\end{eqnarray}
and neglected the squared shear term in the integration and the Doppler terms (see Sec.~\ref{sec2_B} for the validity of the above approximations). From the equation above, we see that the change of the apparent magnitude in an inhomogeneous universe is linearly related  to the perturbation of non-relativistic matter.  

In order to obtain the variance of the apparent magnitude PDF of SNe Ia at the same $z_s$, we need the angular correlation for the apparent magnitude of two sources at points $\hat{\bf{n}}$ and $\hat{\bf{n}}'$ $(\theta \equiv |\hat{\bf{n}} - \hat{\bf{n}}'|)$ over the entire sphere of $z=z_s$. Using Eq.~(\ref{eq.m-z}), we can calculate the angular autocorrelation function in the same way as the derivation of the correlation for convergence~\citep{2001PhR...340..291B}:  
\begin{eqnarray}
	\xi_{\rm lens}(z_s,\theta) 
		&\equiv& \langle \delta m_{\rm lens}(z_s,{\hat{\bf{n}}})\delta m_{\rm lens}(z_s,\hat{{\bf{n}}}') \rangle
		\nonumber \\
		&=& \left(\frac{15H_0^2\Omega_{m0}}{2\ln10}\right)^2 \int_{0}^{\chi_{s}} d\chi \left[\frac{(\chi_s-\chi)\chi}{\chi_s} \Bigl(1+z(\chi)\Bigr)\right]^2
		\nonumber \\
		&& \times \int_0^\infty \frac{d\ln k}{2\pi}k^2 P_{\rm nl}(\chi,k)  J_{0} [\chi \theta k],  \label{am_cor}
\end{eqnarray}  
where $P_{\rm nl}(\chi(z),k)$ is the nonlinear matter power spectrum and $J_{0}(x)$ is the zeroth-order Bessel function. Here we have used the well-established approximation, {\it limber's equation}~\citep{1954ApJ...119..655L,1992ApJ...388..272K}, which relates the angular correlation of the projected field to that of the three-dimensional field. Consequently, the variance of apparent magnitude due to lensing is obtained by taking $\theta = 0$:
\begin{eqnarray}
	\sigma^2_{\rm lens}(z_s) = \langle \delta m_{\rm lens}^2(z_s) \rangle = \xi_{\rm lens}(z_s,0).  \label{sig_lens} 
\end{eqnarray}

\subsection{\label{sec2_B}Approach to small-scale structures}

Another important approach to constraining neutrino masses is {\it cosmic shear} observation, in fact, \citet{2009A&A...500..657T} set, for the first time, a limit on the total mass of neutrinos from CFHTLS cosmic shear data. In the context of comic shear, we focus on the angular correlation of the averaged ellipticity of galaxies within finite areas separated with an angle, $\theta$, which is the same expression as Eq.~(\ref{am_cor}) except for the coefficient~\citep{2001PhR...340..291B}. Therefore, the structure corresponding to a smaller scale than $\theta$ does not affect on the result because of the Bessel function. On the other hand, SNe are point sources, and light rays from SNe are also influenced by smaller-scale structures (we have to deal with the angular autocorrelation at zero lag for this reason), therefore, we can obtain the information on smaller scales. 

However, this effect gives rise to some problems at the same time. First, some of the observed SNe Ia will be  
 {\it strongly} magnified by a gravitational lens, and these should not be included in our treatment because we have assumed  that the magnification is small enough in the previous section. Actually,  the strong lensing probability for SNe with $z \gsim 1$ is about $10^{-3}$, and so far only two SNe are observed to be strongly magnified: PS1-10afx (Type Ia), which is 30 times brighter than normal for its distance by the Panoramic Survey Telescope and Rapid Response System 1~\citep{2013ApJ...767..162C, 2014Sci...344..396Q} and SN Refsdal (not a Type Ia), which was discovered to be split into multiple images by the {\it Hubble Space Telescope}~\citep{2015Sci...347.1123K}. 
However, some optical imaging surveys in the next decade will be able to detect strongly lensed SNe in large numbers, in particular, LSST is expected to find an order of  100 strongly lensed SNe, including 50 SNe Ia~\citep{2010MNRAS.405.2579O}. 
Therefore, we have to {\it systematically} exclude the strongly lensed samples.

The second problem is the shear effect, which was assumed to be neglected in the previous section. \citet{1989PhRvD..40.2502F} showed that discussing the light propagation in a universe filled with objects of a certain size, the squared shear term in Eq.~(\ref{eq.d-z}) is small enough compared with the convergence as long as one focuses on structures above galactic scales as lensing objects.\footnote{They also showed that the convergence term in Eq.~(\ref{eq.d-z}) is dominant compared with the Doppler terms for $z \gsim 10^{-1}$ we are considering here~\citep[see][for a detailed discussion of the Doppler terms]{2006PhRvD..73l3526H}. \label{foot_2}\label{foot_2}} However, if the light rays are influenced by structure below galactic scale, this approximation could break down.
We could understand this picture in the context of strong lensing as follows: the Einstein radius for galaxies is smaller than the virial radius. On the other hand, the Einstein radius for the objects with stellar size is larger than the whole size, that is, the stellar objects can have a  considerable influence on light rays passing through area outside of themselves.        

Then, which kind of light rays are strongly magnified or distorted? In the following argument, we assume that lensing objects are only galaxies and do not include clusters (or groups) of galaxies because it is known that only clusters with high concentration parameters can produce strong lensing images and the number of these are much less than 
normal clusters~\citep{2005ApJ...619L.143B,2007ApJ...654..714H,2008ApJ...684..177U,2009MNRAS.392..930O}.  
In the galaxy-galaxy lensing analysis, the mass profile described by a singular isothermal sphere (SIS) model for the mass profile is often used where the Einstein radius $\theta_{\rm E}$ is given by
\begin{eqnarray}
	\theta_{\rm E}=4\pi \left(\frac{\sigma_v}{c}\right)^{2} \frac{D_{ds}}{D_{s}}.  \label{Ein_radi} 	
\end{eqnarray}         
Here, $\sigma_v$ is the one-dimensional velocity dispersion of a lensing galaxy, $D_{ds}$ is the angular diameter distance between the lens and the source, and $D_{s}$ is the angular diameter distance between the observer and the source. 

This SIS density profile, which is equivalent to the flat rotation curves observed for spiral galaxies, also has been shown to be consistent with the total mass profile of early-type galaxies~\citep{2001ApJ...549L..33R,2005ApJ...623..666R,2006ApJ...649..599K,2007ApJ...667..176G}. For a source at angular position ${\boldsymbol \theta}$, the convergence $\kappa$ and the shear $\gamma \equiv |\gamma| {\rm e}^{2i \varphi}$ caused by an SIS  have the same expression:   
\begin{eqnarray}
	\kappa(\theta) = |\gamma(\theta)| = \frac{\theta_{\rm E}}{2 \theta},  \label{con_she} 		
\end{eqnarray}      
where $\theta \equiv |{\boldsymbol \theta}|$. Here, the magnification $\mu$ can be written, in terms of the convergence and shear, by 
\begin{eqnarray}
	\mu = \frac{1}{(1-\kappa)^2 - |\gamma|^2}.  \label{mag} 		
\end{eqnarray} 
Therefore, we expect that we are not suffering from the above two problems as long as we focus on only the light rays with $\theta > \theta_{\rm E}$ for all lensing galaxies along the line of sight.  Then, we introduce a critical radius $\theta_{c} (> \theta_{\rm E})$ to consider the following sample selection for SNe Ia: “we only use such SNe Ia that the centers of the foreground galaxies are not included in the area within the critical radius $\theta_{c}$ of the SNe Ia.” This selection makes it possible to exclude the strongly lensed SNe and safely adopt the approximation, ignoring the shear effect at the same time. The largest Einstein radius, which corresponds to massive early-type galaxies with a velocity dispersion $\sigma_v \sim 300$ $[{\rm km\ s^{-1}}]$, is estimated to be $\theta_{\rm E} \lsim 3''$ using the fact that $D_{ds}/D_{s} \lsim 1$. This suggests that it is required that $\theta_{c}$ is larger than the order of $1''$.

For the SNe Ia in the range of $z_s \sim 1-2$, whose lensing dispersion is comparable to the dispersion in the intrinsic peak magnitude of SNe Ia, the lensing efficiency becomes the maximum for the lensing galaxies at $z \sim 0.5$. Hence, it follows that the lensing galaxies at $z \sim 0.5$ have the biggest effect on the net lensing dispersion. The order of angular size, at $z=0.5$, for typical galaxy halos, which have the virial masses $M_{\rm vir} \lsim 10^{11} - 10^{12} \ [M_{\odot}]$ and the virial radii $r_{\rm vir} \lsim 10 -10^{2} \ [\rm kpc]$, is estimated to be $r_{\rm vir}/d^{\rm{FRW}}_{\rm{A}} \lsim 1-10\ ['']$, where $d^{\rm{FRW}}_{\rm{A}}$ is the angular diameter distance in a homogeneous FRW universe. In fact, the angular diameter distance, $d^{\rm{FRW}}_{\rm{A}}$, is the same order over the range of $z \gsim 0.1$. Therefore, this estimation for the angular size could be valid for most galaxies along the line of sight. The above discussions mean that there is a critical mass $M_{\rm c}$ that corresponds to the critical radius $\theta_{c}$. That is, the light rays for selected SNe Ia with a critical radius $\theta_{c} \simeq 1-10\ ['']$ are not affected by structures with the smaller masses than the critical mass, $M_{\rm vir} < M_{\rm c} \simeq 10^{11} - 10^{12} \ [M_{\odot}]$. Therefore, we need to connect the correspondence between the critical radius and the critical mass to the theoretical formalism in Sec.~\ref{sec2_A}.

\subsection{\label{sec2_C}Critical value of the wave number}

Perturbations at each spatial scale, corresponding to density contrasts as a continuous fluid or gravitationally collapsed objects, contribute to the lensing dispersion through the integration of the nonlinear power spectrum  over the wave number (the inverse of physical scales) in Eq~(\ref{am_cor}) which is derived under the assumption that 
only weak lensing dominates. 
Hence, we should take account of the selection for SNe Ia 
to eliminate the effects from structures with a smaller scale than $M_{\rm c}$ by setting an upper limit of the wave number. 

In the Press-Schechter model~\citep{1974ApJ...187..425P} based on a spherically symmetric collapse of dust, the (physical) number density of halos of masses between $M$ and $M+dM$ at $z$ is
\begin{eqnarray}
	\frac{dn(z;M)}{dM}=\sqrt{\frac{2}{\pi}}\ \frac{\bar{\rho}_m(z)}{M}\ \frac{d\nu}{dM}\ \exp \left(-\frac{\nu^2}{2}\right), \label{P_S}
\end{eqnarray}
where $\bar{\rho}_m(z)$ is the background matter density and $\nu \equiv \delta_{c}/\sigma_{\rm lin}(z;M)$. Here, $\sigma_{\rm lin}$ is the variance of the smoothed linear density contrast, with the top hat window function in a ball of the present spatial size $R$ and the mean mass $M(R)=(4\pi/3)R^3 \bar{\rho}_{m}(0)$, 
\begin{eqnarray}
	\sigma_{\rm lin}^2(z;M(R))
		\equiv \frac{1}{2\pi^2} \int_{0}^{\infty}dk \ k^2 \ \tilde{W}_R^2(k)P_{\rm lin}(z,k). \label{var_lin}
\end{eqnarray}
where $\tilde{W}_R(k)=3(\sin kR -kR \cos kR)/(kR)^3$ and $P_{\rm lin}(z,k)$ is the linear power spectrum. The critical value $\delta_{c}$ is obtained analytically in the Einstein-de Sitter model: $\delta_{c}=1.686$. In flat models containing the cosmological constant, $\delta_{c}$ depends on $\Omega_{m}(z)$, which is the time-dependent matter density parameter, and changes with redshift. However, the time dependence is very weak~\citep{1996MNRAS.282..263E}, therefore, we use the value $\delta_{c}=1.686$ in what follows.

The variance $\sigma_{\rm lin}(z;M)$ evolves in proportion to the linear growth factor with redshift, on the other hand, at a fixed redshift, $\sigma_{\rm lin}(z;M) \to \infty$ as $M \to 0$. Hence, at each redshift $z$, the critical mass, $M_{\rm coll}(z)$, of objects that begin to be formed at the redshift can be defined as 
\begin{eqnarray}
	\sigma_{\rm lin}(z;M_{\rm coll}(z)) = \delta_{c}/\sqrt{2}, \label{coll_con}
\end{eqnarray}
at which the exponent of the suppression factor in Eq.~(\ref{P_S}) is equal to unity. The critical mass corresponds to the critical spatial size, $R_{\rm coll}$, one-to-one: $M_{\rm coll}=(4\pi/3)R_{\rm coll}^3 \bar{\rho}_{m}(0)$, hence, it follows that at a redshift $z$, the matter fluctuation on scales smaller than $R_{\rm coll}(z)$ exists as some collapsed objects. Now, we introduce the critical wave number given by   
\begin{eqnarray}
k_{c}(z) = \left \{
\begin{array}{l}
2 \pi/R_{\rm coll}(z) \quad \mbox{for} \quad M_{\rm c} > M_{\rm coll}(z) \\
2 \pi/R_{\rm c} \quad \  \mbox{for} \quad M_{\rm c} \leq M_{\rm coll}(z) 
\end{array}
\right. \label{k_c}
\end{eqnarray}
where $R_{\rm c}$ is determined such that $M_{\rm c}=(4\pi/3)R_{\rm c}^3 \bar{\rho}_{m}(0)$. Finally, we can estimate the lensing dispersion due to the contributions from density fluctuations as a continuous fluid at a scale larger than $R_{\rm coll}(z)$ and collapsed objects with masses larger than $M_{\rm c}$ by integrating the nonlinear power spectrum over the range from $0$ to $k_{\rm c}(z)$ in Eq~(\ref{am_cor}).

\begin{figure}[t]
  \begin{center}
   \includegraphics[width=80mm,angle=270]{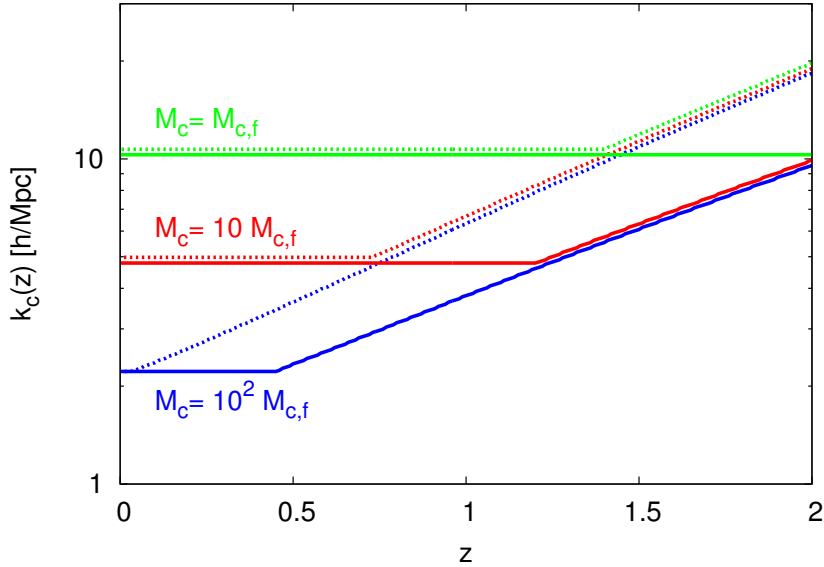}
  \end{center}
  \caption{\label{fig_1} Critical wave number in the fiducial $\Lambda$CDM model (solid lines), compared to a model with three degenerate massive neutrinos, $\Sigma m_{\nu}=0.6\ [{\rm eV}]$, (dotted lines). For each model, $M_{\rm c,f}$, $10 M_{\rm c,f}$ and $10^{2}M_{\rm c,f}$ are shown with green lines, red lines, and blue lines, respectively.}
\end{figure}

\section{\label{sec3}Constraints on neutrino masses}

In this section, we estimate how accurately the total mass of neutrinos is constrained from a given data for the lensing dispersion of SNe Ia. To begin with, we discuss the difference of between $\Lambda$CDM models with massive and massless neutrinos and calculate the lensing dispersion for each model in Sec.~\ref{sec3_A}. After that, in Sec.~\ref{sec3_B}, we introduce the covariance matrix to derive the Fisher matrix, which describes the standard deviation of estimated model parameters, and we forecast constraints on neutrino masses for some future surveys in Sec.~\ref{sec3_C}.

\subsection{\label{sec3_A}$\Lambda$CDM $+$ massive neutrino model}

Taking account of the discussion in Sec~\ref{sec2_B} and \ref{sec2_C}, in this paper, we use the following fiducial value of the critical mass, $M_{\rm c}$, deciding the critical wave number: $M_{{\rm c},f}=10^{11}[M_{\odot}]$. For comparison, we also consider models with $M_{\rm c}=10 M_{{\rm c},f}$ and $M_{\rm c}=10^{2}M_{{\rm c},f}$, which correspond to the larger critical radii. As for the total neutrino mass, our fiducial model (or the (almost) massless model) assumes a normal mass hierarchy with $(\Sigma m_{\nu})_{,f}=0.06[{\rm eV}]$, which means that there is one massive neutrino in the heaviest neutrino mass eigenstate and two massless neutrinos. In addition, we fix the other cosmological parameters to the values presented by the {\it Wilkinson Microwave Anisotropy Probe} five-year release~\citep{2009ApJS..180..330K} unless otherwise noted.

In the linear theory, structure formation in $\Lambda$CDM models including massive neutrinos is characterized by the free-streaming scale, $k_{\rm fs}$, given by 
\begin{eqnarray}
	k_{\rm fs}(z) \simeq \frac{0.677}{(1+z)^{1/2}}\left(\frac{m_{\nu}}{1\ {\rm eV}}\right)(\Omega_{m0}h^2)^{1/2} \ {\rm Mpc^{-1}}
\end{eqnarray}
where $h$ is the Hubble parameter defined as $H_{0}=100 h \ {\rm km\ s^{-1}\ Mpc^{-1}}$~\citep{1980PhRvL..45.1980B,2006PhRvD..73h3520T}. This scale corresponds to the Jeans length, therefore, it follows that neutrino density fluctuations with $k > k_{\rm fs}$ are strongly suppressed and, accordingly, the linear matter power spectrum at $k > k_{\rm fs}$ decreases. Fig.~\ref{fig_1} shows the critical wave number, Eq.~(\ref{k_c}), in the fiducial $\Lambda$CDM model (solid lines) and a model with three degenerate massive neutrinos, $\Sigma m_{\nu}=0.6\ [{\rm eV}]$,\footnote{Although the tightest constraint on the total neutrino mass is $\Sigma m_{\nu} \lsim 0.2\ [{\rm eV}]$ as we mentioned in Sec~\ref{sec1}, we consider this extreme model so that neutrinos' effect on the matter power spectrum or the lensing dispersion of SNe Ia can be seen easily.} (in which we can neglect the small mass splittings) (dotted lines). For each model, $M_{\rm c}=M_{\rm c,f}$, $10 M_{\rm c,f}$, and $10^{2}M_{\rm c,f}$ are shown with green, red, and blue lines, respectively. We find that the massive neutrino model shifts to the side of low redshift compared to the fiducial model since the time that the variance of linear fluctuations defined as Eq.~(\ref{var_lin}) satisfies the condition Eq.~(\ref{coll_con}) is late due to the neutrino suppression of the matter power spectrum. In addition, we can see that the critical wave number has a smaller value as the value of critical mass gets larger, as is clear from Eq.~(\ref{k_c}).

\begin{figure}[t]
  \begin{center}
   \includegraphics[width=80mm,angle=270]{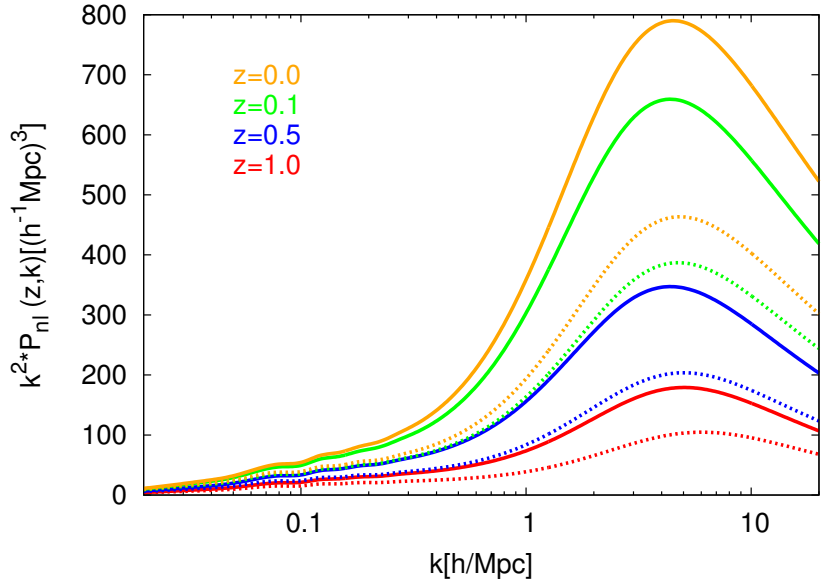}
  \end{center}
  \caption{\label{fig_2} Nonlinear matter power spectrum in the fiducial $\Lambda$CDM model (solid lines) and a model with three degenerate massive neutrinos, $\Sigma m_{\nu}=0.6\ [{\rm eV}]$, (dotted lines). Each color (orange, green, blue, and red) shows the growth with redshift ($z=0.0,0.1,0.5$, and 1.0, respectively).}
\end{figure}

\begin{figure}[t]
  \begin{center}
   \includegraphics[width=80mm,angle=270]{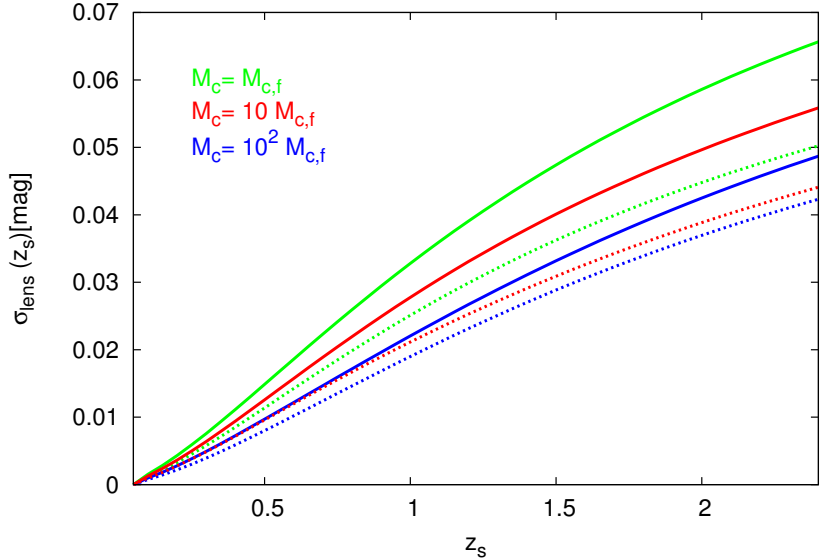}
  \end{center}
  \caption{\label{fig_3} Lensing dispersion of SNe Ia in the fiducial $\Lambda$CDM model (solid lines), compared to a model with three degenerate massive neutrinos, $\Sigma m_{\nu}=0.6\ [{\rm eV}]$, (dotted lines). For each model, $M_{\rm c,f}$, $10 M_{\rm c,f}$ and $10^{2}M_{\rm c,f}$ are shown with green lines, red lines, and blue lines, respectively.}
\end{figure}

It follows from Fig.~\ref{fig_1} that in order to calculate the lensing dispersion for sources at $z\sim1-2$, from Eq.~(\ref{am_cor}), we need the information of the matter power spectrum at $k \sim 10[h{\rm Mpc}^{-1}]$. This scale is the strongly nonlinear regime; therefore,  we have to take it   into consideration in the expression of the nonlinear power spectrum. For example, {\sc halofit}~\citep{2003MNRAS.341.1311S,2012ApJ...761..152T} has been widely used as an accurate fitting formula for the nonlinear matter power spectrum. In this paper, we use an improved fitting formula, proposed by \citet{2012MNRAS.420.2551B}, which modifies {\sc halofit} 
for taking account of the effects of massive neutrinos for $k < 7[h{\rm Mpc}^{-1}]$ at $z \leq 3$. The modified {\sc halofit} confirms that the largest error is roughly 2\% for $\Sigma m_{\nu}=0.3[{\rm eV}]$. In addition, when calculating the linear and nonlinear matter power spectrums, we use {\sc camb}\footnote{\url{http://camb.info/}}~\citep{2000ApJ...538..473L} into which the modified {\sc halofit} has been incorporated. 

We show the nonlinear power spectrum, $k^2 P_{\rm nl}(z,k)$, for the fiducial $\Lambda$CDM model (solid lines) and the model with $\Sigma m_{\nu}=0.6\ [{\rm eV}]$ (dotted lines) in Fig.~\ref{fig_2}. The growth with redshift is represented with colors: $z=$0.0 (orange), 0.1 (green), 0.5 (blue), and 1.0 (red). It is found that the contribution from each scale is maximum at $k \sim 5[h{\rm Mpc}^{-1}]$ and the model with $0.6[{\rm eV}]$ is suppressed to about half in comparison with the fiducial model ($\Sigma m_{\nu}=0.06$) in the scale of $k \sim 1-10[h{\rm Mpc}^{-1}]$. Compared with the order of the critical wave number in Fig.~\ref{fig_1}, it is expected that the lensing dispersion strongly depends on the critical mass as well as the critical radius. 

Fig.~\ref{fig_3} shows the lensing dispersion of SNe Ia, defined as Eq.~\ref{sig_lens}, in the same manner as Fig.~\ref{fig_1}. In the cases with smaller values of the critical mass, which have larger upper limits of the wave number, the lensing dispersions become larger since we also take account of light rays affected by collapsed objects with smaller masses. For models with a certain value of the critical mass, the model with $0.6[{\rm eV}]$ is reduced compared with the fiducial model, hence, it 
follows that the effect of the suppression of the power spectrum is larger than that of the increase in the critical wave number. Furthermore, we can see that the difference between the fiducial model and the model with $0.6[{\rm eV}]$ becomes larger as the critical mass becomes smaller. This means that models with smaller values of the critical mass are more sensitive to the total mass of massive neutrinos. Note that in models with $M_{\rm c}< M_{\rm c,f}$, that is, with $\theta_{c} \lsim 1''$, the shear effect could not be neglected, and strongly magnified SNe Ia could not be excluded as we have discussed in Sec.~\ref{sec2_B}.

\subsection{\label{sec3_B}Fisher matrix analysis}

We have seen how massive neutrinos affect on the lensing dispersion of SNe Ia. Given this, how accurately can we obtain the information about the total mass of massive neutrinos from a given data set? The discussion below basically follows an approach discussed by \citet{1999MNRAS.305..746M}. We consider a vector ${\bf x}$ of a given data set, which consists of $N$ real numbers ($x_{1},\cdots,x_{N})$, and assume its probability distribution $L({\bf x}; {\bf \Theta})$ depends on a vector of $m$ model parameters ${\bf \Theta}=(\theta _{1},\cdots,\theta_{m})$. Then, the Fisher information matrix is defined as  
\begin{eqnarray}
	({\bf F})_{ij} \equiv -\left \langle \frac{\partial^{2} \ln f}{\partial \theta_{i} \partial \theta_{j}} \right \rangle,
\end{eqnarray} 
and its inverse ${\bf F}^{-1}$ gives the standard deviations for the errors on these parameters measured by the maximum likelihood estimate: $\sigma(\theta_{i})=({\bf F}^{-1})_{ii}$, where $\sigma(\theta_{i})$ is the standard deviation of the error on a parameter $\theta_{i}$~\citep[see][for a review]{1997ApJ...480...22T}. For the case in which the PDF $f$ is Gaussian, the Fisher information matrix can be written as  
\begin{eqnarray}
	({\bf F})_{ij} = \frac{1}{2}\ {\rm Tr}\left[{\bf C}^{-1}{\bf C}_{,i}{\bf C}^{-1}{\bf C}_{,j}+{\bf C}^{-1}({\boldsymbol \mu}_{,i}{\boldsymbol \mu}^{t}_{,j}+{\boldsymbol \mu}_{,j}{\boldsymbol \mu}^{t}_{,i})\right], \label{fisher_G}
\end{eqnarray}
where ${\bf C}_{,i}\equiv \partial{\bf C}/\partial \theta_{i}$, ${\boldsymbol \mu}=\langle {\bf x} \rangle$ is the mean vector and ${\bf C}=\langle ({\bf x}-{\boldsymbol \mu})({\bf x}-{\boldsymbol \mu})^{t} \rangle$ is the covariance matrix~\citep{1996ApJ...465...34V}.

In our situation, the observed data vector $x_{i}$ is the set of apparent magnitudes $m_{i}$ of SNe Ia at $z_{i}$, which generally includes not only the lensing dispersion $\delta m_{\rm lens}(z_i)$ but also other random errors with zero mean ($\langle \delta m_{\rm x} \rangle=0 $)~\citep{2014ApJ...780...24S}:
\begin{eqnarray}
	m_{i} = \underbrace{5\log_{10}d_{\rm L}^{\rm FRW}(z_{i}) + M_{i}}_{\mu_{i}} + \delta m_{\rm lens}(z_i) + \delta m_{\rm int} +  \delta m_{\rm fit},  \label{uncertainty}
\end{eqnarray}
where $M_{i}$ is the absolute magnitude of the $i$th SN, $\delta m_{\rm int}$ is the intrinsic dispersion of SNe Ia, and $\delta m_{\rm fit}$ is the uncertainty due to the light-curve fitting in which the photometric measurement error is assumed to be included. Supposing that there are no correlations between different SNe and between different types of errors\footnote{Actually, these correlations could be due to the light-curve fitting model or observational instruments.}, the covariance matrix is    
\begin{eqnarray}
	({\bf C})_{ij} = [\sigma_{\rm lens}^{2}(z_{i}) + \sigma_{\rm c}^{2}]\delta_{ij},
\end{eqnarray}
where $\sigma_{\rm x}^{2} \equiv \langle \delta m_{\rm x}^{2} \rangle$, and $\sigma_{\rm c}^{2} \equiv \sigma_{\rm int}^{2} + \sigma_{\rm fit}^{2}$, which can be recognized as a constant since $\delta m_{\rm int}$ and $\delta m_{\rm fit}$ are essentially not depend on redshift.
Furthermore, we assume that the probability distribution of the net luminosity dispersion, ${\bf x}-{\boldsymbol \mu}$, is Gaussian although at least, the distribution of $\delta m_{\rm lens}$ is not Gaussian because nonlinear growth of the density contrast inevitably gives rise to a skewness of the distribution through Eq.~\ref{eq.m-z}, which has been also suggested by a ray-tracing simulation~\citep{2011ApJ...742...15T}. Though the constant error $\sigma_{\rm c}$ could be determined from observed data of SNe at low redshift where the contribution of the lensing dispersion is negligible, in what follows, we conservatively consider to estimate the total mass of neutrinos and $\sigma_{\rm c}$ at the same time. Therefore, we focus on $\theta_{1}=\Sigma m_{\nu}$ and $\theta_{2}=\sigma_{c}$ as the model parameters and fix the other cosmological parameters, which leads to ${\boldsymbol \mu}_{,i}=0$. Then, it follows from Eq.~\ref{fisher_G} that the Fisher information matrix in our case is given by
\begin{eqnarray}
	({\bf F})_{ij} = \sum_{i=1}^{N} \frac{1}{2}\ \frac{[\sigma_{\rm lens}^{2}(z_{i}) + \sigma_{\rm c}^{2}]_{,i}[\sigma_{\rm lens}^{2}(z_{i}) + \sigma_{\rm c}^{2}]_{,j}}{[\sigma_{\rm lens}^{2}(z_{i}) + \sigma_{\rm c}^{2}]^{2}}. \label{fisher_ourcase}
\end{eqnarray}

\begin{table}[b]
  \begin{center}
	\caption{\label{table_1}Rates of SNe Ia in each $\Delta z = 0.1$ redshift bin for {\it WFIRST} and LSST(main)}
    \begin{tabular}{c|ll} \hline
      Survey  &  {\it WFIRST} \quad  &  LSST(main) \quad \\ \hline 
   $z$=0.2\footnotetext{d}  & $0.6 \times 10^{2}$ &   $4 \times 10^{3}$         \\
	  0.3     & $2.0 \times 10^{2}$ &   $1 \times 10^{4}$         \\
	  0.4     & $4.0 \times 10^{2}$ &   $2 \times 10^{4}$         \\
	  0.5     & $2.2 \times 10^{2}$ &   $1 \times 10^{4}$         \\
	  0.6     & $3.2 \times 10^{2}$ &   $4 \times 10^{3}$         \\
	  0.7     & $1.4 \times 10^{2}$ &   $2 \times 10^{2}$         \\
      0.8-1.7 & $1.4 \times 10^{2}$ (for each bin) &                             \\ \hline
    \end{tabular}
	\vskip 3pt
	\begin{minipage}{460pt}
	Note. We use only SNe Ia at $z>0.2$ to safely ignore the Doppler terms as we mentioned in Sec.~\ref{sec2_B}. 
	\end{minipage}
  \end{center}
\end{table}

\subsection{\label{sec3_C}Forecast}

In order to forecast to what extent we can constrain the total mass of neutrinos from future optical imaging surveys, we use the expected numbers of SNe Ia for the {\it WFIRST}~\citep[][]{2015arXiv150303757S} and the main survey of the LSST \citep[LSST(main);][]{2009arXiv0912.0201L}, which are summarized in Table~\ref{table_1}. Note that, in our forecast, we do not use SNe Ia at at $z<0.1$, where the Doppler terms could be dominant in Eq.~\ref{eq.d-z} (see footnote \ref{foot_2}). Moreover, we consider, as a fiducial value of $\sigma_{c}$, $\sigma_{c,f}=0.11$, which corresponds to the error model adopted, in the {\it WFIRST} project, to forecast the performance of the supernova survey.\footnote{The {\it WFIRST}-AFTA 2015 Report~\citep{2015arXiv150303757S} adopted the error model that the photometric measurement error per SN is $\sigma_{\rm meas}=$0.08[mag] ($\sigma_{\rm fit} \geq \sigma_{\rm meas}$) and the intrinsic luminosity dispersion in Type Ia is $\sigma_{\rm int}=$0.08[mag].} As for the other parameters, we use the fiducial values introduced in Sec.~\ref{sec3_A}: $M_{\rm c,f}=10^{11}[M_{\odot}]$, $(\Sigma m_{\nu})_{,f}=0.06[{\rm eV}]$, etc.

\begin{figure}[t]
\begin{tabular}{cc}
 \begin{minipage}{0.5\hsize}
  \begin{center}
   \includegraphics[width=60mm,angle=270]{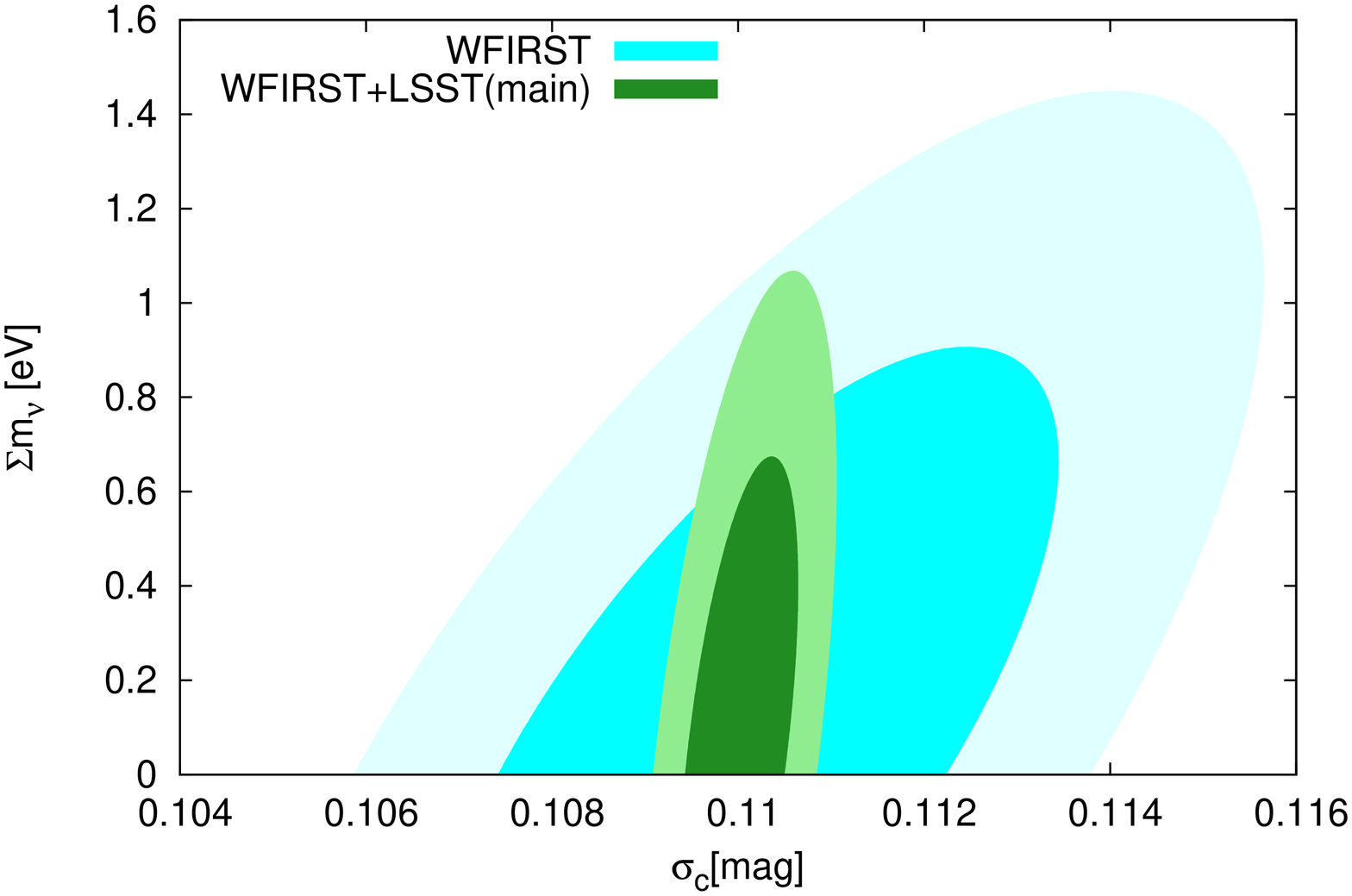}
  \end{center}
 \end{minipage}
 \begin{minipage}{0.5\hsize}
  \begin{center}
   \includegraphics[width=60mm,angle=270]{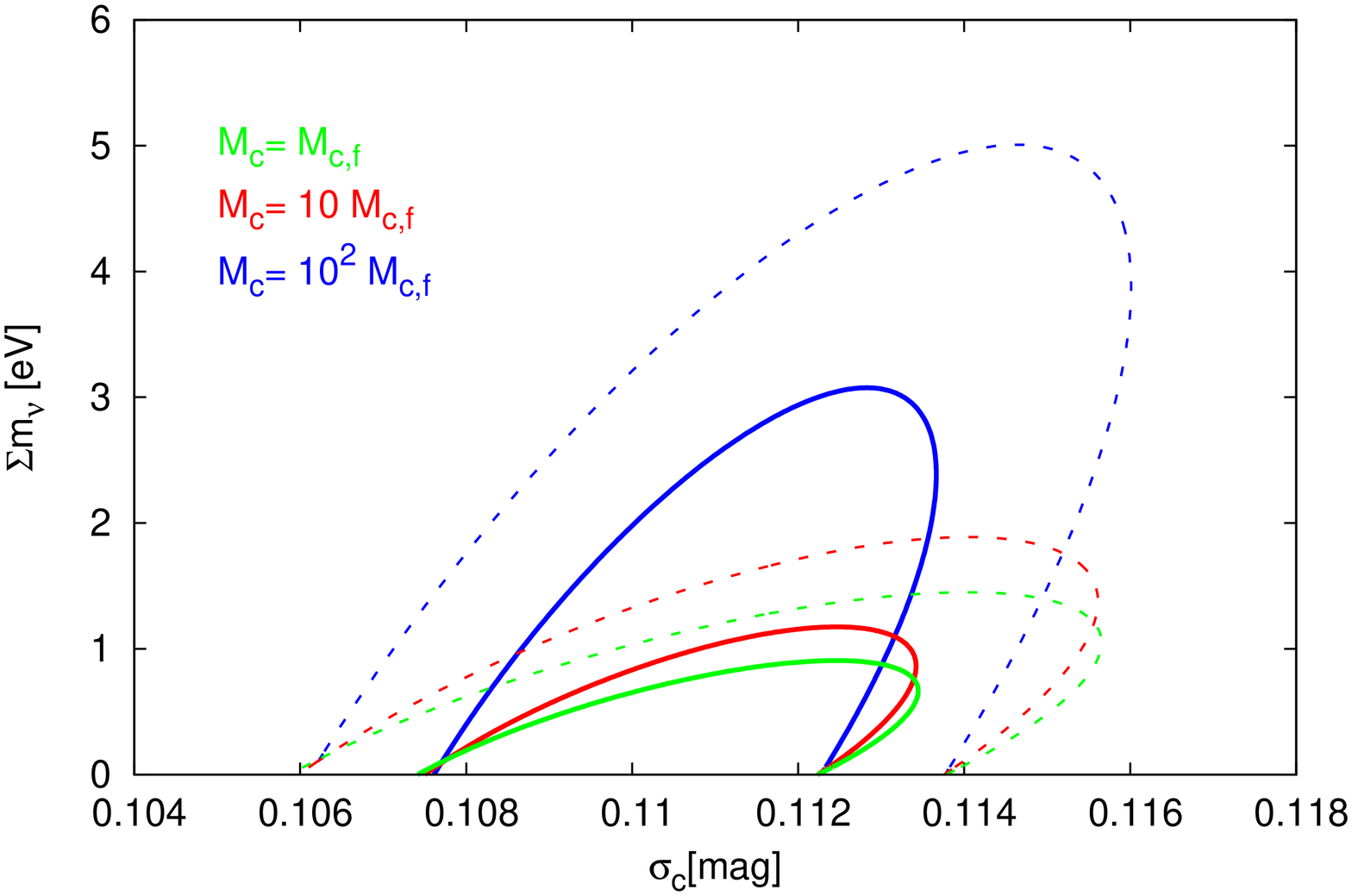}
  \end{center}
\end{minipage}
\end{tabular}
  \caption{\label{fig_4} Forecast of the constraints on neutrino masses from several future surveys. \it Left panel: the contours show $1\sigma$ and $2\sigma$ for our fiducial model. The cyan region is expected from {\it WFIRST} and the green region from both of {\it WFIRST} and LSST(main). Right panel: the solid and dashed lines show $1\sigma$ and $2\sigma$, respectively, expected from {\it WFIRST}. $M_{\rm c,f}$, $10 M_{\rm c,f}$ and $10^{2}M_{\rm c,f}$ are described with green lines, red lines, and blue lines, respectively.}
\end{figure}

The left panel of Fig.~\ref{fig_4} shows $1\sigma$ and $2\sigma$ contours expected from {\it WFIRST} only (cyan region) and from both of {\it WFIRST} and LSST(main) (green region). Note that when calculating the errors on parameters from the combination of {\it WFIRST} and LSST(main), we only use LSST(main) in the range of $z=0.2-0.7$, taking the possibility of data overlap into consideration. First, we see that the constraint on neutrino masses expected from {\it WFIRST} in our fiducial model is $\Sigma m_{\nu}<1.5[{\rm eV}]$(95\% CL). In addition, although the expected number of SNe Ia for LSST(main) is two orders of magnitude greater than that for {\it WFIRST} in the range of $z=0.2-0.7$, the constraint expected from the both is $\Sigma m_{\nu}<1.1[{\rm eV}]$(95\% CL), which is slightly better than for {\it WFIRST} only. This means that SNe Ia at high redshift $z\gsim1$ are sensitive to the total mass of neutrinos, which is consistent with the fact that the difference between the lensing dispersion for models with massive and massless neutrinos increases with redshift (see Fig.~\ref{fig_3}). Furthermore, in the right panel of Fig.~\ref{fig_4}, we show $1\sigma$ (solid lines) and $2\sigma$ (dashed lines) expected from {\it WFIRST} for $M_{\rm c,f}$(green), $10M_{\rm c,f}$(red), and $10^{2}M_{\rm c,f}$(blue). We can find clearly that the constraints for models with a smaller value of the critical mass are better, which reflects the discussion in Sec.~\ref{sec3_A} that these models are more sensitive to neutrino masses.

\begin{figure}[t]
\begin{tabular}{cc}
 \begin{minipage}{0.5\hsize}
  \begin{center}
   \includegraphics[width=60mm,angle=270]{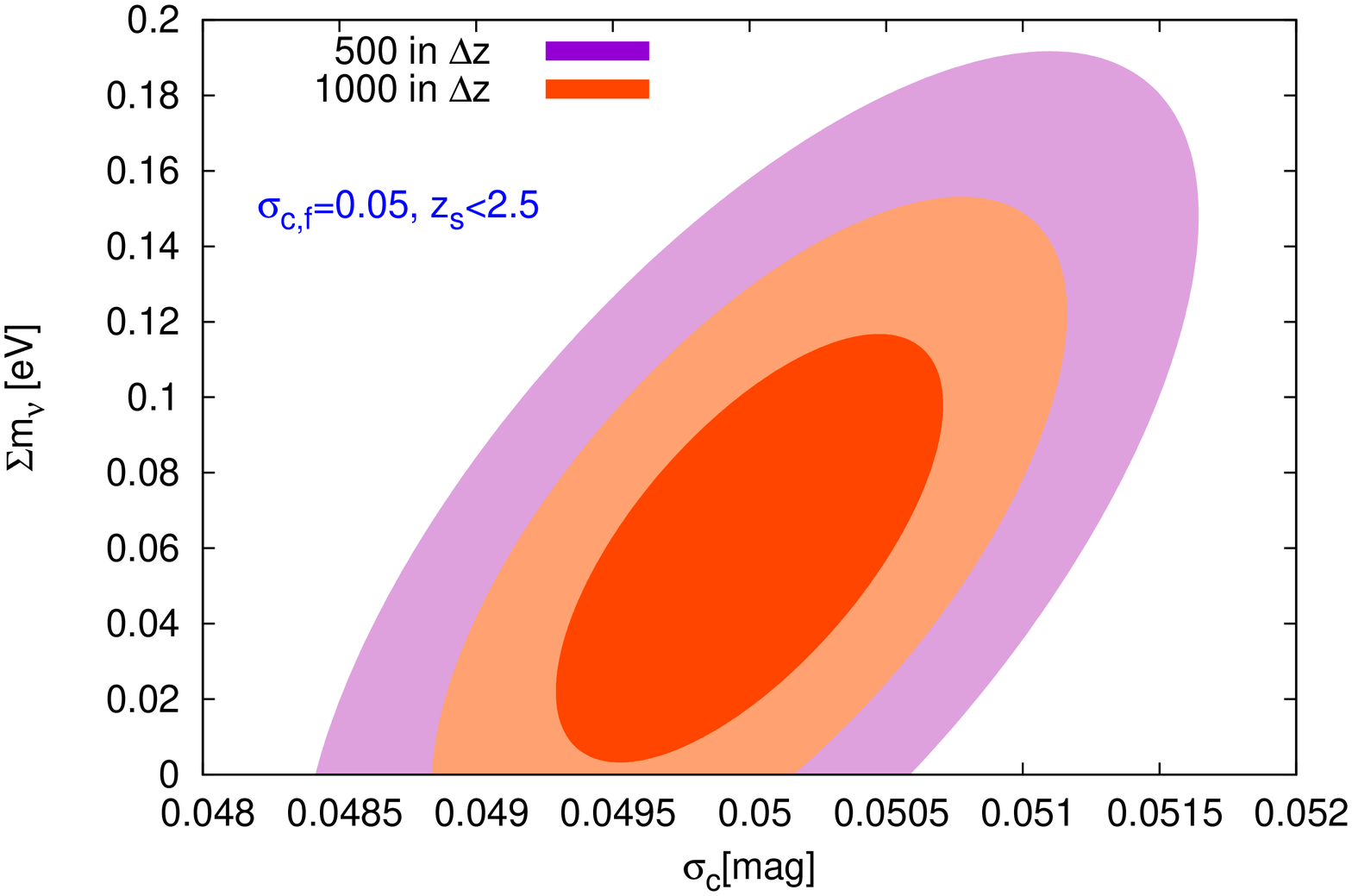}
  \end{center}
 \end{minipage}
 \begin{minipage}{0.5\hsize}
  \begin{center}
   \includegraphics[width=60mm,angle=270]{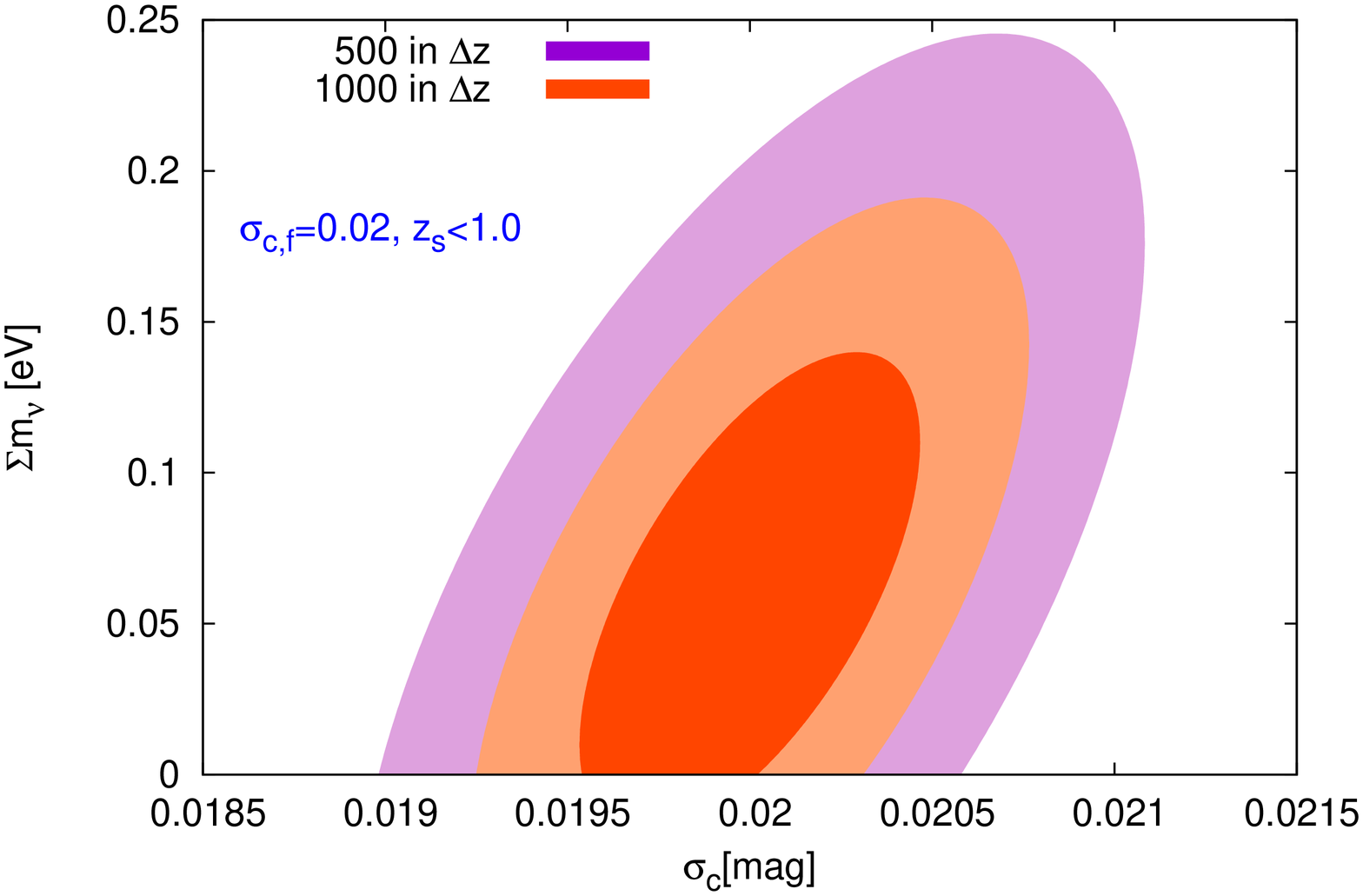}
  \end{center}
\end{minipage}
\end{tabular}
  \caption{\label{fig_5} Source redshifts and value of the constant error $\sigma_{\rm c}$ required toward stronger constraints on the total mass of neutrinos. Left panel: the contours show $1\sigma$ and $2\sigma$ in the model with $\sigma_{\rm c,f}=0.05$, $z_{s}<2.5$. It shows the two cases: $500$ (purple region) or $1000$ (orange region) SNe Ia in each $\Delta z = 0.1$ redshift bin. Right panel: similarly, in the model with $\sigma_{\rm c,f}=0.02$, $z_{s}<1.0$.}
\end{figure}

These results suggests that it is difficult to constrain the total mass of  neutrinos beyond the strongest constraint, $\Sigma m_{\nu} \lsim 0.2[{\rm eV}]$, using the lensing dispersion of SNe Ia from future optical imaging surveys. How far will we have to observe SNe Ia to surpass the best limit? To what extent will we have to reduce the constant error $\sigma_{c}$ not caused by lensing? The contours in the left and right panels of Fig.~\ref{fig_5} correspond to $1\sigma$ and $2\sigma$ limits for the following two models: $\sigma_{\rm c,f}=0.05$, $z_{s}<2.5$ and $\sigma_{\rm c,f}=0.02$, $z_{s}<1.0$, respectively. For each model, we assume that we observe $500$ (purple region) or $1000$ (orange region) SNe Ia in each $\Delta z = 0.1$ redshift bin. It follows that we could obtain the constraint on neutrino masses, $\Sigma m_{\nu}\lsim 0.2[{\rm eV}]$(95\% CL), from both models. In other words, even if we can observe only a number of SNe Ia in each redshift bin smaller by an order of magnitude than the expected number for LSST(main), the combination of deepness of survey $z\gsim2.5$ and smallness of the constant error, about half of our fiducial value ($\sigma_{c,f}=0.11[{\rm mag}]$), makes it possible to constrain neutrino masses beyond the strongest constraint. Furthermore, these results also mean that if we can reduce the constant error to $\sigma_{c} = 0.02[{\rm mag}]$, even surveys covering only up to $z \sim 1$ will reach the best limit. In addition, we emphasize that the large amount of SNe Ia at $z\lsim0.7$ expected from LSST(main) is inefficient for constraining neutrino masses; however, is expected to play a important role in reducing the constant error.

\section{\label{sec4}Summary and discussion}

In this paper, we have estimated how accurately the total mass of neutrinos is constrained using the lensing dispersion of SNe Ia. First, we have introduced the relation between the magnification (or demagnification) and the matter density contrast along a line of sight and then obtained the expression of variance of apparent magnitude PDF of SNe Ia due to lensing. Furthermore, we have discussed about a selection of SNe Ia to avoid some difficulties caused by small-scale structures, such as strong lensing or shear effect, and introduced a critical value of the wave number in order to connect the selection to the theoretical formulation. Subsequently, we have  calculated the lensing dispersion for $\Lambda$CDM models with massive and massless neutrinos and forecast the error on the parameter $\Sigma m_{\nu}$ from some future optical imaging surveys using the Fisher information matrix. Finally, we have found that the constraint expected from both {\it WFIRST} and LSST(main) in our model is $\Sigma m_{\nu}<1.1[{\rm eV}]$(95\% CL). We also need the data of SNe Ia from deeper surveys, $z_{s}\gsim2.5$, and must reduce the constant error except for the lensing dispersion to about half of our fiducial value $\sigma_{c,f}=0.11[{\rm mag}]$ (which is adopted to forecast the performance of the SNe survey in {\it WFIRST}) in order to reach the current tightest limit, $\Sigma m_{\nu} \lsim 0.2[{\rm eV}]$.

We note that there are some uncertainties in our model. First, the selection criterion $\theta_{c}$ proposed in Sec.~\ref{sec2_B} have to be verified or improved because we simply treat all lensing objects as SIS and not all SNe surveys are sufficiently deep to image the intervening lensing galaxies. We need to make sure that strongly magnified SNe are not included in samples chosen by this criterion, otherwise we must use the larger value for $\theta_{c}$. The second problem is caused by the treatment of the effect of discrete collapsed objects by introducing a critical wave number. The halo mass function Eq.~(\ref{P_S}) is extremely broad and accordingly the critical wavenumber defined through Eq.~(\ref{coll_con}) has some uncertainty. Therefore this definition should be tested and calibrated from ray-tracing simulations, etc. Furthermore, we discussed the relation between the critical mass $M_{\rm c}$ and the associated critical radius $\theta_{c}$ to select the SNe Ia sample, however, of course this correspondence has to be adjusted as well as the above caliblations because it was based on an order estimation. Note that the right panel of Fig.~\ref{fig_4} suggests that models with smaller values of the critical mass give better constraints as we discussed in Sec.~\ref{sec3_C}, however, a critical mass smaller than $10^{11}\  [M_{\odot}]$ could make it impossible to ignore the shear effects discussed in Sec.~\ref{sec2_B}.

The formalism constructed in this paper holds, if not for light rays, for null geodesics. For example, supermassive binary black holes (BBHs) known as sources of gravitational waves, are potentially powerful standard candles (1\% accuracy for determination of distance: $\delta_{d, {\rm int}} \sim 0.01$) if the location on the sky and redshift are independently determined by an electromagnetic counterpart~\citep{2005ApJ...629...15H}. Thus it is expected that we can obtain the information of LSS using the lensing magnification for gravitational waves in the same way as for SNe Ia. Moreover, the Einstein telescope, which is a next-generation gravitational wave detector, is expected to reach $z \sim 1$  for gravitational waves from BBH with a total mass larger than $10 [M_{\odot}]$~\citep{2012CQGra..29l4013S}. Therefore, taking account of the fact that it is estimated, in this case, that $\sigma_{c} = (5/\ln10) \delta_{d, {\rm int}} \sim  0.02$ (see Eq.~(\ref{eq.m-z})), we find that the right panel of Fig.~\ref{fig_5} shows that gravitational waves from BBHs have the possibility to constrain neutrino masses enough to reach the current best limit. It is difficult to obtain the intrinsically different information from SNe Ia and BBH as standard candles because of the same formalism, however, BBH is still a very interesting event as well as SNe Ia in that they could complement each other.  

In addition, although we have focused on the variance of magnitude PDF in this paper, in principle, we can calculate the higher statistics of PDF and skewness from Eq.~(\ref{eq.m-z}). In Eq.~(\ref{uncertainty}) in Sec.~\ref{sec3_B}, we considered some factors causing the magnitude dispersion (around the mean value $\mu_{i}$). In fact, since only $\delta m_{\rm lens}$ has a non-Gaussian distribution, if we can detect the skewness of the magnitude PDF from SNe observations, that would mean that only the contribution due to lensing is taken out. That is, we do not suffer from the size of the constant error, unlike with the variance. Note that, when computing the skewness, we require the theoretical model for the bispectrum that can reproduce the nonlinearity of the density contrast up to the scale corresponding to the critical wave number, $ k \sim 10[h{\rm Mpc}^{-1}]$.

\acknowledgments

We would like to thank T. Yamada and N.S. Sugiyama for useful discussions. This work is supported in part by JSPS Research Fellowships for Young Scientists (No. 16J01773 to R.H.) and a Grant-in-Aid for Scientific Research from JSPS (No. 26400264 to T.F.).

\bibliographystyle{apj.bst}
\bibliography{ms.bib}

\end{document}